\documentclass[twocolumn,final,twoside,journal]{IEEEtran}
\usepackage{epsfig,amsfonts}
\usepackage[nolist]{acronym}
\usepackage{graphicx,amssymb,amsmath,amsthm}
\usepackage[nospace,noadjust]{cite}
\usepackage{color}
\usepackage{psfrag}
\usepackage{mathtools}
\usepackage{placeins}
\usepackage{xspace}
\usepackage{flushend}

\begin{acronym}%[nolist] % usage: \ac{SW}, \acp{SW} for plurals \acf{SW} Use the full name of the acronym. \acs{SW}Use the acronym, even before the first corresponding \ac command
\acro{RV}{random variable}
\acro{MUD}{Multi User Detection}
\acro{AWGN}{Additive White Gaussian Noise}
\acro{IC}{Interference Cancelation}
\acro{SIC}{Successive Interference Cancelation}
\acro{E-SSA}{Enhanced Spread Spectrum Aloha}
\acro{SSA}{Spread Spectrum Aloha}
\acro{pdf}{probability density function}
\acro{cdf}{cumulative distribution function}
\acro{pmf}{probability mass function}
\acro{PER}{Packet Error Rate}
\acro{SNIR}{Signal to Noise and Interference Ratio}
\acro{i.i.d.}{independent and identically distributed}
\acro{CDMA}{Code Division Multiple Access}
\end{acronym}

\AtBeginDocument{ \renewcommand{\d}{\ensuremath{\mathrm{d}}} }

\newcommand{\sslen}{n}
\newcommand{\K}{k}
\AtBeginDocument{ \renewcommand{\L}{l} }

\renewcommand{\P}{P}
\newcommand{\p}{p}

\newcommand{\crosscorr}{\Psi}

\newcommand{\z}{{q}}
\newcommand{\Z}{{Q}}

\newcommand{\zz}{{u}}
\newcommand{\ZZ}{{U}}

\newcommand{\Ex}{\ensuremath{\mathop{{}\mathbf{E}}}\xspace}
\newcommand{\Var}{\ensuremath{\mathop{{}\mathrm{Var}}}\xspace}

\newcommand{\snirtarget}{\gamma_{\text{t}}}

\newcommand{\snireff}{\Gamma}

\newcommand{\pmin}{p_{\min}}
\newcommand{\pmax}{p_{\max}}

\newcommand{\pmindb}{\theta_{\min}}

\newcommand{\pmaxdb}{\theta_{\max}}

\newcommand{\ica}{\alpha}
\newcommand{\icb}{\delta}

\newcommand{\h}{\mathsf {h}}
\newcommand{\g}{\mathsf {g}}

\newtheorem{prop}{Proposition}

\newcommand{\prodlog}{\mathtt{W}}

\mathtoolsset{showonlyrefs}

\title{Decentralized Power Control for Slotted Spread Spectrum Aloha with Successive Interference Cancellation }
\author{Francisco L\'azaro\\
 Institute of Communications and Navigation, DLR (German Aerospace Center), We\ss ling, Germany\\
Email: {\tt Francisco.LazaroBlasco@dlr.de}%, Markus.Stinner@tum.de}
\thanks{This work has been accepted for publication at the 11th International {ITG} Conference on Systems, Communications and Coding, SCC 2017.}
\thanks{\copyright 2016 IEEE. Personal use of this material is permitted. Permission
from IEEE must be obtained for all other uses, in any current or future media, including
reprinting /republishing this material for advertising or promotional purposes, creating new
collective works, for resale or redistribution to servers or lists, or reuse of any copyrighted
component of this work in other works}
}

\begin{document}
\maketitle
% \date{\today}
% \thispagestyle{empty}
% \setcounter{page}{0}

%%%%%%%%%%%%%%%%%%%%%%%%%%%%%%%%%%%%%%%%%%%%%%%%%%%%%%%%%%%%%%%%%%%%%%%%%%%%%%%%%%%%%%%%%%%%%%%%%%%%%%%%%%

\begin{abstract}
In this paper, we study slotted \acl{SSA} with \acl{SIC} at the receiver over a Gaussian channel. We consider a decentralized power control setting in which each
user chooses its transmit power independently at random according to a power distribution with continuous support. In this setting, we derive an analytical expression for
the expected interference power experienced by a user. This allows us to derive analytically the power distribution that, during the \acl{SIC} process leads to a constant signal to noise plus interference ratio for all users. We consider both perfect and imperfect interference cancellation.
\end{abstract}
%\vspace{-1pt}

% {\pagestyle{plain} \pagenumbering{arabic}}
%%%%%%%%%%%%%%%%%%%%%%%%%%%%%%%%%%%%%%%%%%%%%%%%%%%%%%%%%%%%%%%%%%%%%%%%%%%%%%%%%%%%%%%%%%%%%%%%%%%%%%%%%%
\section{Introduction}\label{sec:Intro}
In many wireless communication settings, such as in satellite communications, it is not possible to use contention  avoidance mechanisms in random access channels.
Traditional wireless systems usually employ the well known Aloha or Slotted Aloha \cite{Abramson:ALOHA,Abramson77:PacketBroadcasting} protocols for random access and operate in a regime in which packet collisions are destructive.  In this setting, the spectral efficiency of random access channels is usually low. Thus, whenever possible on demand resource assignment has traditionally been preferred for data transmission, especially when the amount of data to be transmitted is large.

However, this trend has changed in the last few years mainly due to the rise of machine-to-machine applications, characterized by large populations of terminals transmitting small messages in a bursty manner.
In these conditions on demand resource assignment is known to be inefficient, mainly due to the signalling overhead involved. In order to make machine-to-machine communications more efficient, the research community has proposed a series of random access protocols that leverage mostly on interference cancellation and allow achieving a high spectral efficiency, see e.g. \cite{DeGaudenzi09:CRDSA, delrio:12, Liva2011:IRSA_TCOM}.

\ac{E-SSA} was introduced in \cite{delrio:12} and has been adopted in several standards, e.g., {S-MIM}, \cite{ETSI_TS_102_721}.
\ac{E-SSA} is also employed by the Eutelsat Broadcast Interactive System to provide interactive TV services and machine-to-machine connectivity, whose terminals are known as \emph{Smart LNBs} \cite{fsim}.
In \ac{E-SSA}, the transmitters operate like in an Aloha system but use direct sequence spread spectrum.
At the receiver side, packet oriented \acl{SIC} is used.
More precisely, the receiver stores the received waveform and whenever a packet is successfully decoded, it cancels the interference created by the packet in the stored received signal.
The performance of \ac{E-SSA} depends strongly on the distribution of the power with which the packets arrive at the receiver. Concretely, it has been observed that the throughput of \ac{E-SSA} can increase when there is power unbalance among users \cite{delrio:12}, \cite{rossi:2015}.

Power control for spread spectrum communications in the context of demand assigned multiple access is a very well studied topic.
Indeed, already in  1990, Viterbi \cite{54460} studied the problem of transmission power allocation for a deterministic number users employing equal transmission rates and \ac{CDMA} in the case in which the receiver uses \ac{SIC}.
Viterbi proved that the optimal power allocation was such that if one orders the users according to their transmission power, the power of the $j$-th user grows exponentially with $j$.

In the context of random access, one does not know which users are active, and therefore, it is not possible to control the transmit power of each user individually. Hence, power control is implemented by specifying a transmit power probability distribution. Each active user then generates its transmit power independently at random according to this probability distribution.
In \cite{xu:13} the problem of decentralized power allocation in random access systems was analyzed, and it was found out that when the user population is finite, the support of the power distribution is also finite. Furthermore, the case of a system with two users was studied in depth providing the optimal power distribution, whereas for larger populations of users only a suboptimal power distribution was proposed. In \cite{lin:15} a decentralized power allocation approach is proposed for systems with finite user populations, which also leads to a power distribution with finite support.
In \cite{Collard:14} a semi-analytical approach was followed to determine the optimal power distribution for \ac{E-SSA} with \ac{SIC} at the receiver. Concretely, the optimal power distribution leads to  a uniform distribution of the transmit power in a logarithmic scale (in dB), which is inline with the result obtained by Viterbi in \cite{54460} for demand assigned multiple access.

In this paper, we focus on a simplified setting inspired by \ac{E-SSA}. Concretely we consider slotted spread spectrum Aloha with \ac{SIC} at the receiver and chip synchronous users. Furthermore, we assume that each user employs a different random spreading sequence. These assumptions are made in order to make the analysis tractable.
Furthermore, we make the assumption that the user population is large, since this is usually the case in typical \ac{E-SSA} applications. Hence, in contrast to \cite{xu:13,lin:15} we work with continuous power distributions.
In contrast to \cite{Collard:14} we follow a fully analytical approach, which allows us to derive power distributions fulfilling a design criterium not only for the case of perfect \ac{SIC} but also for the case in which the interference cancellation is imperfect. In the case of perfect \ac{SIC} our results are inline with the results in \cite{Collard:14}.

The remaining of the paper is structured as follows.
In Section~\ref{sec:sys_model}, we introduce the system model. Section~\ref{sec:interf_analysis} presents an analysis of the multiuser interference in the system under the assumption of perfect interference cancellation.
In Section~\ref{sec:opt_dist} we derive the power distribution that leads to a constant signal to noise plus interference ratio for all users during the \ac{SIC} process.
Section~\ref{sec:imperfect} then extends the results to the case in which the interference cancellation is imperfect.
Finally, Section~\ref{sec:conc} contains the conclusions to the work and a discussion about possible extensions and the applicability of the obtained results in practice.

%%%%%%%%%%%%%%%%%%%%%%%%%%%%%%%%%%%%%%%%%%%%%%%%%%%%%%%%%%%%%%%%%%%%%%%%%%%%%%%%%%%%%%%%%%%%%%%%%%%%%%%%%%
\section{System Model}\label{sec:sys_model}
Consider an uplink scenario where a large population of users employs direct sequence spread spectrum slotted Aloha in order to transmit data to a single receiver.
For simplicity, we assume an \ac{AWGN} channel. All users employ  BPSK modulation and different random spreading sequences of length $\sslen$ chips, whose length coincides with the symbol duration.
Thus, the spreading factor of the system corresponds to $\sslen$. All packets are assumed to have the same  length, $\L$ symbols, and users are assumed to be chip and slot synchronous.
For simplicity, we assume that the number of users transmitting a data packet in every slot is constant\footnote{In a real system, the load can be accurately modeled as a Poisson \ac{RV}. Results for a Poisson distributed load can be easily derived from the results for constant load.} and we denote it by $\K$.
Furthermore, we assume that every time a user transmits a packet, the user selects its transmit power $\p$ independently at random according to a \acf{pdf} $f_{\P}(\p)$ with continuous support. Under these assumptions the received signal is
\begin{align}
y(t) =  \sum_{i=1}^{\K}  \sum_{h=1}^{\L}\sqrt p_i \, e^{j \phi_i}\, b_i[h] \, s_i (t - hT ) \,+ \, w(t)
\label{eq:sys_model}
\end{align}
where
\begin{itemize}
  \item $T$ is the symbol interval
  \item $p_i$ is the transmit power used by the $i$-th user to transmit its packet
  \item $\phi_i \in [0, 2 \pi)$ is the phase of user $i$.
  \item $b_i[h] \in  \{-1, +1\} $ is the $h$-th modulated symbol transmitted by user $i$ (BPSK modulation is assumed)
  \item $s_i$ is the signature (spreading) waveform  of user $i$, that is normalized to have unit energy
  \item$w(t)$ is the additive white Gaussian noise.% with density $N_0/2$. %and per dimension variance $\sigma^2/2$
\end{itemize}

At the receiver side, single-user matched filtering\footnote{In theory other detection techniques could be used. However, implementing more advanced detection techniques is challenging in practice, mainly because obtaining channel state information is challenging in a random access setting. The use of linear minimum mean square error detectors for \ac{E-SSA} has been investigated in \cite{gallinaro:2015} and \cite{rossi:2015}.} and packet-wise \ac{SIC} are used.
Thus, the receiver stores the received waveform of the slot of interest in a memory. We assume perfect channel state information at the receiver, and for simplicity we assume the receiver orders the users according to their transmit power, so that ${\p_1 \leq \p_2 \leq  \hdots \leq  \p_\K}$.
The receiver first attempts to decode the user with the highest power, $\p_\K$. If the packet of the $\K$-th user is successfully decoded, its waveform is reconstructed and cancelled from the stored waveform. The receiver then tries to decode the next strongest user, the $\K-1$-th user, and the process is repeated until all users are decoded or until decoding of one of the user fails. Thus, we assume that when the $i$-th user cannot be decoded, users 1 to $i-1$ cannot be decoded either.

%For simplicity we assume that a user can be decoded with probability $1$ whenever its signal to noise plus interference ratio, $\snir$ is above a threshold value $\snirth$, and unsuccessful otherwise (with probability $1$).

%In this paper, we use uppercase letters to denote \acfp{RV}, and lowercase letters to denote their realizations. For example, $x$  denotes a realization of the \ac{RV} $X$.
%%%%%%%%%%%%%%%%%%%%%%%%%%%%%%%%%%%%%%%%%%%%%%%%%%%%%%%%%%%%%%%%%%%%%%%%%%%%%%%%%%%%%%%%%%%%%%%%%%%%%%%%%%
\section{Interference Analysis}\label{sec:interf_analysis}
%In this section, we analyze the  performance of the slotted spread spectrum aloha system described in Section~\ref{sec:sys_model} under the assumptions that the length of the spreading sequence $\sslen$ is large, yet finite, and that the system load $\beta$ is moderate or high. It follows implicitly that the number of active users $\K$ is finite, yet large.

For the analysis, we focus on the packet from a randomly chosen user, user $m+1$ who has transmit power $\P=\p$. Let us also make the assumption that all users with power higher than $\p$ have been successfully decoded and their interference has been cancelled completely. Under these assumptions, the sample associated to the $h$-th transmitted bit at the output of the $m+1$-th user matched filter is:
\begin{equation}
y_{m+1} (t) =  \sqrt p \, b_{m+1}[h] + \sum_{i=1}^{m}  \sqrt p_i \, e^{j \phi_i}\,  \crosscorr_{i,m+1} \, b_i[h] \,+ \, w_{m+1}[h]
\label{eq:su_mf_out}
\end{equation}
where
\begin{itemize}
\item $m$ is the number of users with  transmission power lower than $\p$
\item $\crosscorr_{i, m+1}$ is the crosscorrelation coefficient between the signatures of users $i$ and $m+1$
\item  $w_{m+1}[h]$ is the filtered Gaussian noise, with variance given by $\sigma_w^2$
\end{itemize}
and for notational simplicity we assumed $\phi_{m+1}=0$.

The first term in \eqref{eq:su_mf_out} corresponds to the useful signal and the second term corresponds to multi-user interference. If we denote the interference by $z$ we have
\begin{align}
z &=\sum_{i=1}^{m} z_i = \sum_{i=1}^{m}  \sqrt p_i \, e^{j \phi_i}\, \crosscorr_{i, m+1} \, b_i[h]
\label{eq:z_def}
\end{align}
where
\begin{equation}
z_i = \sqrt p_i \, e^{j \phi_i} \, \crosscorr_{i, m+1} \, b_i[h]
\label{eq:z_i_def}
\end{equation}
is the interference caused by the $i$-th user to the $m+1$-th user.

Let us denote by $\sigma_z^2$ the second moment of the interference suffered by a user with transmit power $p$,
\[
\sigma_z^2:=\Ex[z^2].
\]
The expression for $\sigma_z^2$ is given in the following proposition
\begin{prop}
The total interference power experienced by a user with $\P=\p$ is given by
\[
\sigma_z^2 =  \frac{1}{2} \beta  \int_{0}^{\p} { \p f_{\P}(\p) \, \d \p}
\]
where $\beta$ will be referred to as system load and is given by
\[
\beta = \frac{k-1}{n}
\]
\end{prop}
\begin{IEEEproof}
Let us start by deriving the interference power from each individual user, $\Ex[{z_i}^2]$.  From \eqref{eq:z_i_def} we have that $z_i$ is a product of different terms that have a random nature.

We shall first consider $\crosscorr_{i, m+1}$, the signature crosscorrelation between a generic interferer $i$ and the user of interest $m+1$. Since users are  chip synchronous and use (different) random spreading sequences, the first two moments of $\crosscorr_{i, m+1}$ are (see \cite{verdu1998multiuser}, Chapter~2)
\begin{align}
 &\Ex [\crosscorr_{i, m+1}] =  0  \\
 &\Ex [\crosscorr^2_{i, m+1}] =  \frac{1}{\sslen}.
\end{align}

If we now bring into consideration the random phase shift among users, we have (\cite{verdu1998multiuser}, Chapter~2)
\begin{align}
 &\Ex [e^{j \phi_i}\,\crosscorr_{i, m+1}] =  0  \\
 &\Ex [(e^{j \phi_i}\,\crosscorr_{i, m+1})^2] =  \frac{1}{2 \sslen}.
\end{align}

Let us now consider the transmission power of the interferers,  $p_i$. Recall that each user selects its transmission power $\p_i$ independently at random and according to a distribution $f_\P(\p)$, and also that we assume that all interferers with $p_i>p$ have been decoded and their interference cancelled. Let us introduce random variable $\Z$ to refer to the transmission power of our interferer $p_i$. For for $\z<p$ we have
\begin{align}
f_{\Z}(\z) &= f_{\P}(\z| \z< \p)=  \frac{f_{\P}(\z)}{\int_{0}^{\p}{f_{\P}(\p) \, \d \p}} =\frac{f_{\P}(\z)}{{F_\P(p)}}
\label{eq:pdf_z}
\end{align}
where $F_\P(p)$ is the \ac{cdf} of random variable $P$,
\begin{align}
F_\P(p) := \int_{0}^{\p}{f_{\P}(\p) \, \d \p}.
\label{eq:F_p}
\end{align}
The mean transmission power of an interferer, $\mu_{\Z}$ is
\begin{align}
\mu_{\Z} &= \mathbf{E} [ \Z ] = \int_{0}^{\p} { \z f_{\Z}(\z) \, \d \z} \\
&= \frac{1}{F_\P(p)} \int_{0}^{\p} { \p f_{\P}(\p) \, \d \p}
\end{align}

We can now derive the moments of $(\sqrt p_i e^{j \phi_i}\,\crosscorr_{i, m+1})$, obviously, the first moment is
\begin{align}
 \Ex [\sqrt p_i \, e^{j \phi_i}\,\crosscorr_{i, m+1}] =  0
\end{align}
and the second moment is
\begin{align}
 \Ex [(\sqrt p_i \, e^{j \phi_i}\,\crosscorr_{i, m+1})^2] = \Ex[p_i] \,  \Ex [(\, e^{j \phi_i}\,\crosscorr_{i, m+1})^2] = \mu_\Z \frac{1}{2 \sslen}
\end{align}

Now we are in the position of computing the moments of $z_i$.  The first moment is, of course $0$, $\Ex[z_i]=0$, whereas the second moment is given by
\begin{align}
\Ex[{z_i}^2] = \Ex [(\sqrt p_i \, e^{j \phi_i}\,\crosscorr_{i, m+1})^2 (b_i[h])^2] = \mu_\Z \frac{1}{2 \sslen}
\label{eq:var_z}
\end{align}
since $\Ex [(b_i[h])^2 ]=1$ (BPSK modulation). %\miss{do we need bpsk modulation as assumption?, can we relax it?}

We are now in the position to calculate the total interference power, $\sigma_z^2$. Looking at \eqref{eq:z_def} we can see how $z$ is a sum of $m$ \ac{i.i.d.} random variables. However, the number of interferers, $m$ is also a random variable since users choose their transmission power independently at random.  Hence, $z$ is a random sum of \ac{i.i.d.} random variables.

Let us denote by $M$ the random variable associated to $m$, the number of effective interferers of our user of interest (the $m+1$-th user), who has power $\P=\p$.
Since the transmit powers of the users are \ac{i.i.d.} and distributed according to $f_\P(\p)$, $M$ conditioned to $\P=\p$ is binomially distributed with
\begin{align}
 \Pr(M=m| \P=\p) = \binom {\K-1} {m} F_\P(p)^m (1-F_\P(p))^{\K-1-m}
\end{align}
where $F_\P(p)$ is the cumulative distribution function of the random variable $P$, i.e., the probability that a user transmits with power lower or equal than $\p$, given in \eqref{eq:F_p}.

Finally, we can obtain $\sigma_z^2$ as
\begin{align}
\sigma_z^2 &= \Ex\left[ \left( \sum_{i=1}^{M} z_i \right)^2 \right] = \Var \left[  \sum_{i=1}^{M} z_i \right] \\
&=  \Ex [ M ] \, \Var [ z_i]  +  \, \left( \Ex [ z_i ] \,\right)^2  \Var [M ]  \\
&=   \,(\K-1) \, F_\P(p) \, \Ex[{z_i}^2]  =\,(\K-1) \, F_\P(p) \,  \mu_\Z \frac{1}{2 \sslen}\\
&=  \frac{1}{2} \beta  \int_{0}^{\p} { \p f_{\P}(\p) \, \d \p}
  \label{eq:sigmaz}
\end{align}
where $\beta = (k-1)/n$.

\end{IEEEproof}

%%%%%%%%%%%%%%%%%%%%%%%%%%%%%%%%%%%%%%%%%%%%%%%%%%%%%%%%%%%%%%%%%%%%%%%%%%%%%%%%%%%%%%%%%%%%%%%%%%%%%%%%%%
\section{Power Distribution}\label{sec:opt_dist}

Under the assumption that all users with transmission power $\P>\p$ have been decoded and their interference cancelled, we define the signal to noise plus interference ratio, $\snireff$ of a user with transmission power $\P=\p$ by
\begin{align}
\snireff= \frac {\p}{\sigma_w^2 + \sigma_z^2 }
\label{eq:snr_effective}
\end{align}
where $\sigma_w^2$ is the noise power (its variance at the output of the single user matched filter).

In the considered setting, it seems reasonable to choose a power distribution $f_{\P}(\p)$ so that all users, independently of the transmit power $\p$ they select, experience the same signal to noise plus interference ratio, $\snireff$. Formally, we set the constraint
\begin{equation}
\snireff = \snirtarget
\label{eq:snir_eq_target}
\end{equation}
for all transmit powers $\p$ in the support of $f_{\P}(\p)$ (where $f_{\P}(\p)>0$).  The target signal to noise ratio, $\snirtarget$, can be regarded as a design parameter. In a practical system, $\snirtarget$ should correspond to a signal to noise and interference ratio that allows decoding successfully with high probability with the modulation and coding scheme used.

If we develop equations \eqref{eq:snr_effective} and \eqref{eq:snir_eq_target}  we have
\begin{align}
\frac{\p}{\sigma_w^2 + \frac{1}{2} \beta  \int_{0}^{\p} { \p f_{\P}(\p) \, \d \p } }&= \snirtarget \\
 \snirtarget \left( \sigma_w^2 + \frac{1}{2} \beta  \int_{0}^{\p}  \p f_{\P}(\p) \, \d \p  \right) &= \p.
\end{align}

In order to solve this integral equation we can simply take the derivative with respect to $\p$ at both sides of the equation, obtaining
\begin{equation}
\frac{1}{2} \snirtarget \, \beta \, \p f_{\P}(\p)= 1.
\end{equation}
Hence, we have
\begin{align}
 f_\P(\p) =  \frac{2}{\snirtarget \,  \beta} \frac{1}{\p}.
\end{align}

Let us denote as $\pmin$ and $\pmax$ respectively the minimum and maximum transmit power. According to our design criterium, we would like to have $\snireff(p) = \snirtarget$ for all transmit powers. Since $f_\P(\p)$ is continuous, a user transmitting with $\pmin$ will suffer no interference, hence we will have
\[
\pmin = \snirtarget \, \sigma_w^2.
\]
We shall assume that $\pmax$ is given, for example due to the limitation of the amplifier in the terminals. Hence, the value of the system load $\beta$  for which we will have ${\snireff = \snirtarget}$ is given by:
\[
\beta = 2 \frac{1}{\snirtarget} \log \left( \frac{\pmax}{\pmin}\right).
\]
Thus, the system load $\beta$ that can be supported (while keeping $\snireff$ constant) grows logarithmically with the maximum transmit power of the terminals $\pmax$.

The power distribution can be recast in terms of $\pmin$ and $\pmax$ as
\begin{align}
 f_\P(\p) =  \begin{cases} \frac{1}{ \log \left( \frac{\pmax}{\pmin}\right) } \frac{1}{\p}, & \pmin\leq \p  \leq \pmax \\
 0,   & \text{otherwise.} \end{cases} \label{eq:opt_dist}
\end{align}

If we introduce in \eqref{eq:opt_dist} the variable change
\[
{\theta = \h(\p) =  10 \log_{10} \p}
\]
we obtain the power distribution with the power expressed in dB.
The function $\h(\p)$ is strictly increasing, continuous and differentiable and has inverse $\p = \g(\theta)= 10^{\theta/10}$. Thus, the \ac{pdf} of $\theta$ can be obtained as
\begin{equation}
  f_\Theta(\theta) = f_\P( g(\theta) ) g'(\theta)
  \label{eq:var_change}
\end{equation}

which leads to,
\begin{align}
 f_\Theta(\theta) =  \begin{cases} \frac{1}{ \pmaxdb - \pmindb},  &  \pmindb < \theta < \pmaxdb \\
 0,   & \text{otherwise} \end{cases}
\end{align}
where $\pmaxdb$ and $\pmindb$ are respectively $\pmax$ and $\pmin$ expressed in dB,
\begin{align}
\pmaxdb &= 10 \log_{10} \pmax \\
\pmindb &= 10 \log_{10} \pmin.
\end{align}
Thus, expressed in dB, the power distribution obtained is uniform, which is in line with the result obtained in \cite{Collard:14} for \ac{E-SSA} using a semi-analytical approach.

%\miss{check what an exponential power profile is}
%\begin{clar}
%The authors in \cite{Collard:14} also state that the optimal power distribution leads to an exponential power profile, where by power profile it is meant the quantile function. The quantile function of random variable $\P$ is given by the value at which the $\Pr(\P \leq p)$ is equal to $\p$.
%
%From \eqref{eq:opt_dist} it is easy to obtain the \ac{cdf} as
%\[
%F_\P(\p)= \int_{\pmin}^{\p} f_\P(\p) \d \p  = \frac{1}{ \log \left( \frac{\pmax}{\pmin}\right) } \log\left(\frac{p}{\pmin}\right) .
%\]
%The quantile function $Q_\P(\p)$ is obtained by setting $ F_\P( Q_\P(\p) )=p$, which leads to
%\[
%Q_\P(\p) = \pmax e^\p
%\]
%for $\pmin \leq \p \leq \pmax$.
%\end{clar}

%%%%%%%%%%%%%%%%%%%%%%%%%%%%%%%%%%%%%%%%%%%%%%%%%%%%%%%%%%%%%%%%%%%%%%%%%%%%%%%%%%%%%%%%%%%%%%%%%%%%%%%%%%
\section{Imperfect Interference Cancellation}\label{sec:imperfect}
In this section we consider the case in which the interference cancellation is imperfect. More concretely we consider two different models of imperfect interference cancellation and derive analytically in both cases the power distribution that yields to a constant signal to noise and interference ratio $\snireff=\snirtarget$ for all values of $\p$.

\subsection{Constant Interference Cancellation Efficiency}\label{sec:imperfect_a}
Let us focus again on the output of the $m+1$-th  matched filter output, which has a transmission power $\P=\p$. Again, we assume that users with transmission power $\P>\p$ have been successfully decoded, but now we assume their interference  has not been completely cancelled. Concretely, we assume that a fraction $1-\ica$ of the interference power has been cancelled, and a fraction $\ica$ is still present. Hence, at the output of the $m+1$-th user matched filter we have
\begin{align}
y_{m+1} (t) &=  \sqrt p \, b_{m+1}[h] + \sum_{i=1}^{m}  \sqrt p_i \, e^{j \phi_i}\,  \crosscorr_{i, m+1} \, b_i[h] \\
& + \sum_{i=m+2}^{k}  \sqrt \ica \sqrt p_i \, e^{j \phi_i}\,  \crosscorr_{i, m+1} \, b_i[h] + \, w_{m+1}[h]
\label{eq:su_mf_out_imperf_a}
\end{align}

This expression can be recast as

\begin{equation}
y_{m+1} (t) =  \sqrt p \, b_{m+1}[h] + \sum_{i=1}^{k-1}  x_i \,+ \, w_{m+1}[h]
\label{eq:su_mf_out_imperf_a1}
\end{equation}
where $x_i$ is the interference caused by the $i$-th interferer,  we have $k-1$ interferers, and the interferers are \ac{i.i.d.}. Concretely, we have that for each interferer with probability ${\Pr(\P \leq p)= F_\P(p)}$ the interferer has transmit power lower than $\p$, and hence its transmit power is distributed according to \eqref{eq:pdf_z}. Thus, we have
\begin{align}
\Ex[{x_i}^2| \P \leq p] = \mu_\Z \frac{1}{2 \sslen}.
\label{eq:var_z_a_1}
\end{align}
Note this is exactly the case we considered in the previous section (see \eqref{eq:var_z}).

In the complementary case, with probability ${\Pr(\P > p)= 1-F_\P(p)}$  the interferer has power higher than $\p$. Denoting by $\ZZ$ the random variable associated to the interference power of users with transmit power $p_i>p$, we have
\begin{equation}
f_{\ZZ}(\zz) = f_{\P}(\zz| \zz > \p)=  \frac{f_{\P}(\zz)}{\int_{\p}^{\pmax}{f_{\P}(\p) \, \d \p}} =\frac{f_{\P}(\zz)}{{1- F_\P(p)}}.
\label{eq:pdf_u}
\end{equation}
According to our assumptions, interferers with transmit power higher than $\p$ have already been  decoded and their interference partially canceled. More concretely, a fraction $\ica$ of their power is still present after interference cancellation, leading to
\begin{align}
\Ex[{x_i}^2| \P>  p] = \ica \frac{1}{2 \sslen} \frac{1}{ 1-F_\P(p)} \int_{p}^{\infty} \p f_\P (p) \d \p
\label{eq:var_z_a_2}
\end{align}

Hence, we have
\begin{align}
\Ex[{x_i}^2] &=  \Ex[{x_i}^2| \P\leq  p]  \Pr(\P \leq p) + \Ex[{x_i}^2| \P>  p]  \Pr(\P > p) \\
&= \frac{1}{2 \sslen} \left( \int_{0}^{\p}{ \p f_{\P}(\p) \, \d \p }+ \ica \int_{\p}^{\infty}{ \p f_{\P}(\p) \, \d \p} \right).
\label{eq:var_z_a_3}
\end{align}
Thus, due to independence among the different $x_i$ the total interference power becomes
\[
\sigma_z^2 := \Ex \left[ \left( \sum_{i=1}^{k-1} x_i \right)^2 \right] = (k-1) \Ex[{x_i}^2]
\]
with $\Ex[{x_i}^2]$ given in \eqref{eq:var_z_a_3}.

Again we design the power distribution in order to obtain ${\snireff = \snirtarget}$, which yields
\begin{equation}
 f_\P(\p) =  \frac{2}{\snirtarget \,  \beta} \frac{1}{\p} \frac{1}{1-\ica}.
 \label{eq:pdf_imperf_a}
\end{equation}

If we now assume $\pmax$ is given we obtain
\begin{align}
  \pmin = \frac{\snirtarget \left( \sigma_w^2 + \frac{\ica}{1-\ica} \pmax \right)}{1 + \snirtarget  \frac{\ica}{1-\ica} }
\end{align}
and the load $\beta$ for which we obtain ${\snireff = \snirtarget}, \forall \p$ is
\begin{align}
  \beta =  \frac{1}{1-\ica} \frac{2}{\snirtarget} \log \left( \frac{\pmax}{\pmin}\right).
\end{align}
In the logarithmic domain (in dB) the distribution becomes again a uniform distribution.

\begin{figure}
  \centering
  \includegraphics[width=\columnwidth]{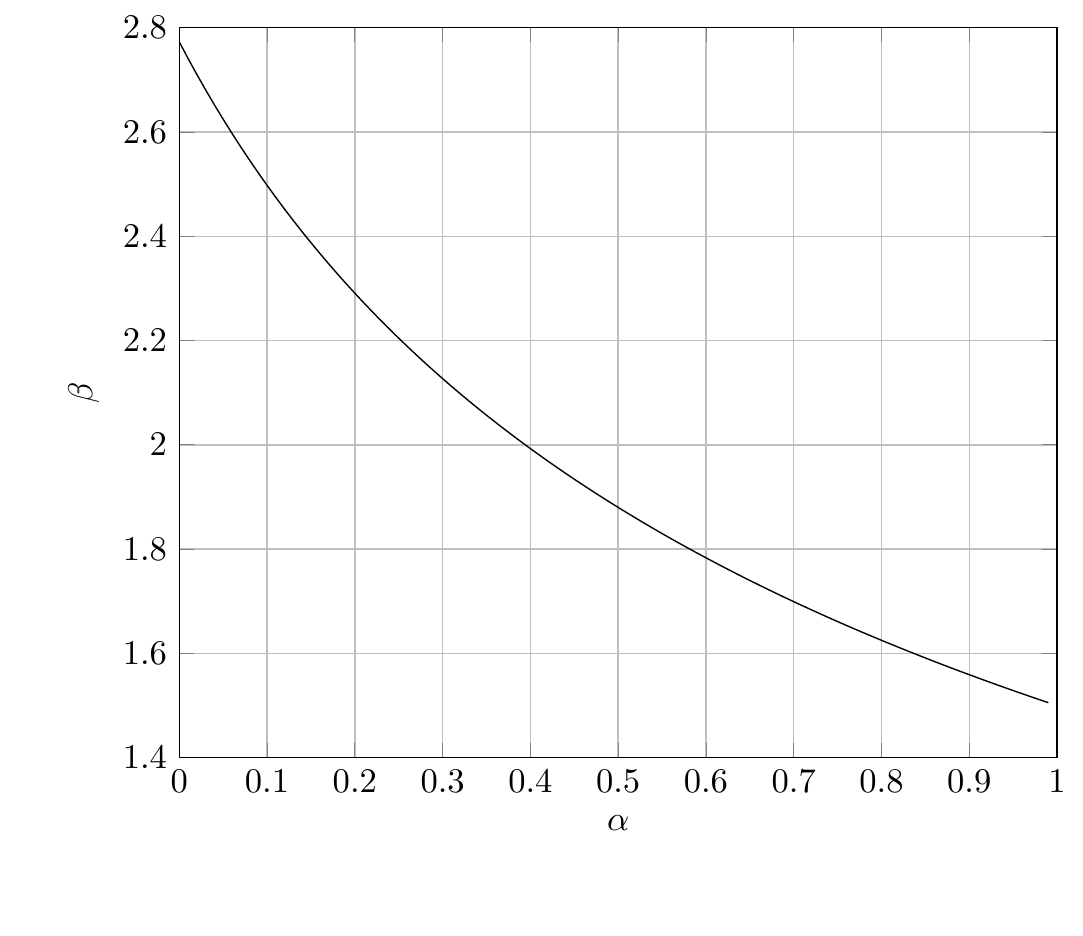}
  \caption{System load $\beta$ as a function of $\ica$,  the fraction of remaining interference power after interference cancellation. A system with $\sigma_w^2  = 1$, $\snirtarget =1$ and $\pmax=4$ is considered.}
  \label{fig:beta_imperfect_1}
\end{figure}

For illustration, in Fig.~\ref{fig:beta_imperfect_1} we provide a numerical example for a system with ${\sigma_w^2  = 1}$, ${\snirtarget =1}$ and ${\pmax=4}$, in which users select their transmit power according to \eqref{eq:pdf_imperf_a}. Concretely, the figure shows the dependence of the system load $\beta$ on $\ica$, the fraction of remaining interference power. We can see how $\beta$ decreases as $\ica$ increases. Note that for $\ica=0$ we have perfect interference cancellation whereas for $\ica=1$ no interference cancellation is taking place.

\subsection{Constant Remaining Interference Power}\label{sec:imperfect_b}
Here we assume that  when interference cancellation is carried out there is a constant interference power $\icb$ that is not cancelled. Hence, the output of the $m+1$-th user output filter becomes
\begin{align}
y_{m+1} (t) &=  \sqrt p \, b_{m+1}[h] + \sum_{i=1}^{m}  \sqrt p_i \, e^{j \phi_i}\,  \crosscorr_{i, m+1} \, b_i[h] \\
& + \sum_{i=m+2}^{k}  \sqrt \icb \, e^{j \phi_i}\,  \crosscorr_{i, m+1} \, b_i[h] + \, w_{m+1}[h]
\label{eq:su_mf_out_imperf_b}
\end{align}
where we assumed again that the $m+1$-th user has transmission power $p$.

We recast this expression as
\begin{equation}
y_{m+1} (t) =  \sqrt p \, b_{m+1}[h] + \sum_{i=1}^{k-1}  x_i \,+ \, w_{m+1}[h]
\label{eq:su_mf_out_imperf_b1}
\end{equation}
where $x_i$ is the interference caused by the $i$-th interferer, and we have $k-1$ interferers that are \ac{i.i.d.}. For each interferer we have that with probability $\Pr(\P \leq p)= F_\P(p)$ the interferer has power lower than $\p$. Hence its power distribution is given by \eqref{eq:pdf_u} and its second moment is given in \eqref{eq:var_z}.

In the complementary case, we have that with probability ${\Pr(\P > p)= 1-F_\P(p)}$  the (remaining) interference has amplitude $\sqrt \icb$ and its second moment corresponds to
\begin{align}
\Ex[{x_i}^2| \P>  p] = \icb \frac{1}{2 \sslen}.
\label{eq:var_z_b_2}
\end{align}

Thus, we have
\begin{align}
\Ex[{x_i}^2] &=  \Ex[{x_i}^2| \P\leq  p]  \Pr(\P \leq p) + \Ex[{x_i}^2| \P>  p]  \Pr(\P > p) \\
&= \frac{1}{2 \sslen} \left( \int_{0}^{\p}{ \p f_{\P}(\p) \, \d \p }+ \icb \left( 1- F_\P(\p)\right) \right)
\label{eq:var_z_b_3}
\end{align}
and the total interference power becomes
\[
\sigma_z^2 := \Ex \left[ \left( \sum_{i=1}^{k-1} x_i \right)^2 \right] = (k-1) \Ex[{x_i}^2]
\]
with $\Ex[{x_i}^2]$ given in \eqref{eq:var_z_b_3}.

We design the power distribution to obtain ${\snireff = \snirtarget}$, which leads us to
\begin{equation}
 f_\P(\p) =  \frac{2}{\snirtarget \,  \beta} \frac{1}{\p-\icb}.
 \label{eq:pdf_imperf_b}
\end{equation}

If we assume $\pmax$ is given we can obtain $\pmin$ and $\beta$ by solving these two equations
\begin{align}
&\pmin = \snirtarget \left( \sigma_w^2 + \icb \frac{\beta}{2} \right)\\
&\beta =  \frac{2}{\snirtarget} \log \left( \frac{\pmax - \icb}{\pmin - \icb} \right)
\end{align}
which are obtained respectively by setting $\snireff = \snirtarget$ for $\p=\pmin$ and by imposing $F_\P(\pmax)=1$. This yields:
\begin{equation}
\pmin = \icb \, \prodlog \left(\frac{e^{\frac{\snirtarget \sigma_w^2}{ \icb} \,-1 } \, (\pmax - \icb) } {\icb} +1 \right)
\end{equation}
for $\icb \leq \pmax$ and $\snirtarget \leq \pmax / \sigma_w^2$,
and where $\prodlog$ is the Lambert W function, also known as product logarithm function.

By introducing the variable change in \eqref{eq:var_change} we obtain the power distribution in dB,
\begin{align}
 f_\Theta(\theta) =   \frac{2}{ \snirtarget \beta} \frac{\log 10}{10} \frac{10^{\theta/10}}{10^{\theta/10}-\icb}
 \end{align}
for $\pmindb < \theta < \pmaxdb$. We can observe how when $\icb=0$ we have a uniform distribution. However, if $\icb>0$ the distribution is not uniform.

\begin{figure}
  \centering
  \includegraphics[width=\columnwidth]{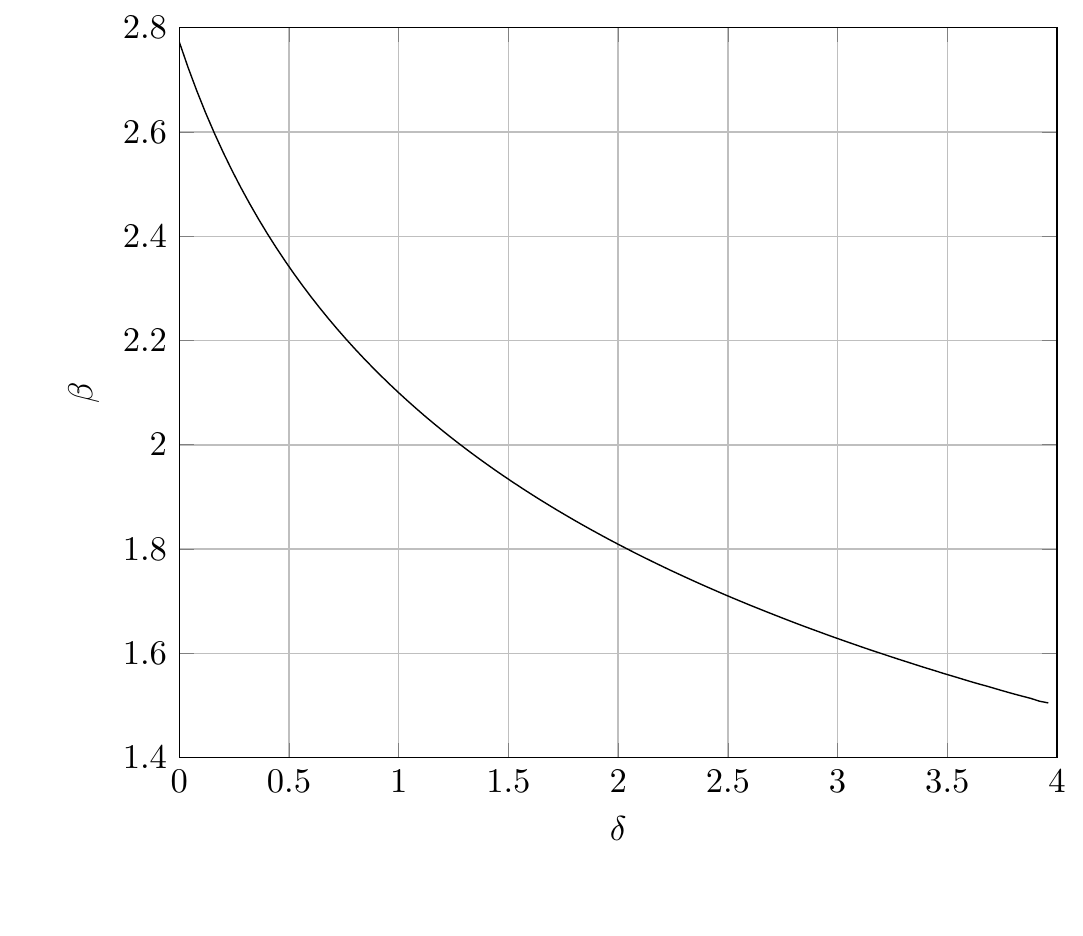}
  \caption{System load $\beta$ as a function of $\icb$, the remaining interference power after interference cancellation. A system with $\sigma_w^2  = 1$, $\snirtarget =1$ and $\pmax=4$ is considered.}
  \label{fig:beta_imperfect_2}
\end{figure}

In order to illustrate the impact of $\icb$ we provide a numerical example in Fig.~\ref{fig:beta_imperfect_2}. We consider a system with ${\sigma_w^2  = 1}$, ${\snirtarget =1}$ and ${\pmax=4}$, in which users select their transmit power according to \eqref{eq:pdf_imperf_b}. The figure shows the dependence of the system load $\beta$ on $\icb$, the remaining interference power after interference cancellation. We can observe how $\beta$ decreases as $\icb$ increases. We remark that for ${\icb=0}$ we have perfect interference cancellation whereas for ${\icb=\pmax}$ no interference cancellation is taking place.
%%%%%%%%%%%%%%%%%%%%%%%%%%%%%%%%%%%%%%%%%%%%%%%%%%%%%%%%%%%%%%%%%%%%%%%%%%%%%%%%%%%%%%%%%%%%%%%%%%%%%%%%%%
\section{Conclusions and Discussion}\label{sec:conc}
In this paper we have studied slotted \acl{SSA} over a Gaussian channel with \ac{SIC} at the receiver  and decentralized power control. Concretely, we have considered that each user chooses its transmit power independently at random according to a power distribution with continuous support. For this setting we derive first the analytical expression for
the expected interference power experienced by a user as a function of its transmit power.  Based on this result we then derive analytically the power distribution that has the property of leading to a  constant signal to noise plus interference ratio for all users. Although initially perfect interference cancellation is assumed, the results are then extended to imperfect interference cancellation. Concretely two different models of imperfect \ac{SIC} are considered.

The setting considered in this paper presents several differences compared to a typical \ac{E-SSA} setting. Concretely, we considered chip and slot synchronous users, whereas users are totally asynchronous in \ac{E-SSA}. Furthermore, we assumed that users employ independent random spreading sequences, whereas in \ac{E-SSA} typically all users employ the same spreading sequence. The extension of the present work to chip, bit and slot asynchronous users is left for further work. Nevertheless, the author expects the results obtained in this paper to hold for \ac{E-SSA} as a first order approximation.

%%%%%%%%%%%%%%%%%%%%%%%%%%%%%%%%%%%%%%%%%%%%%%%%%%%%%%%%%%%%%%%%%%%%%%%%%%%%%%%%%%%%%%%%%%%%%%%%%%%%%%%%%%
\section{Acknowledgements}
The author would like to thank Gianluigi Liva and Markus Stinner for the fruitful discussions.

The research leading to these results has been carried out under the framework of the project `R\&D for the maritime safety and security and corresponding real time services'. The project started in January 2013 and is led by the Program Coordination Defence and Security Research within the German Aerospace Center (DLR).
%%%%%%%%%%%%%%%%%%%%%%%%%%%%%%%%%%%%%%%%%%%%%%%%%%%%%%%%%%%%%%%%%%%%%%%%%%%%%%%%%%%%%%%%%%%%%%%%%%%%%%%%%%
%%%%% Literature
\bibliographystyle{IEEEtran}
\bibliography{IEEEabrv,studio3}

\end{document}